\documentclass[sigconf]{acmart}

\setcopyright{none}
\renewcommand\footnotetextcopyrightpermission[1]{}

\acmConference[EASE 2026]
{The 30th International Conference on Evaluation and Assessment in Software Engineering}
{June 9--12, 2026}
{Glasgow, Scotland, United Kingdom}

\usepackage{placeins}
\usepackage{graphicx}
\usepackage{multirow}
\usepackage{array}
\usepackage{microtype}
\usepackage{xcolor}
\usepackage{enumitem}
\usepackage{booktabs}

\title{Context Before Code: An Experience Report on Vibe Coding in Practice}

\author{Md Nasir Uddin Shuvo}
\email{nasir.shuvo@tuni.fi}
\affiliation{
  \institution{Faculty of Information Technology and Communication Sciences, Tampere University}
  \city{Tampere}
  \country{Finland}
}

\author{Md Aidul Islam}
\email{mdaidul.islam@tuni.fi}
\affiliation{
  \institution{Faculty of Information Technology and Communication Sciences, Tampere University}
  \city{Tampere}
  \country{Finland}
}

\author{Md Mahade Hasan}
\email{mdmahade.hasan@tuni.fi}
\affiliation{
  \institution{Faculty of Information Technology and Communication Sciences, Tampere University}
  \city{Tampere}
  \country{Finland}
}

\author{Muhammad Waseem}
\email{muhammad.waseem@tuni.fi}
\affiliation{
  \institution{Faculty of Information Technology and Communication Sciences, Tampere University}
  \city{Tampere}
  \country{Finland}
}

\author{Pekka Abrahamsson}
\email{pekka.abrahamsson@tuni.fi}
\affiliation{
  \institution{Faculty of Information Technology and Communication Sciences, Tampere University}
  \city{Tampere}
  \country{Finland}
}

\renewcommand{\shortauthors}{Shuvo et al.}

\begin{document}

\begin{abstract}
Code-generating tools are increasingly used in software development, yet experience reports on conversational “vibe coding” under production constraints remain limited. This paper presents an experience report from a small full-stack team that applied contextual prompting and explicit architectural constraints to build (i) a multi-project agent learning platform designed for sustained, production-oriented use and (ii) an academic retrieval-augmented generation system.

The agent platform supports multiple isolated projects, each with structured memory and background processing, thereby enforces project-level isolation. The RAG system provides citation-grounded answers, role-based access control, and evaluation tracking.

Across both systems, vibe coding accelerated scaffolding and integration. However, the generated code often under-specified isolation rules and infrastructure constraints when these were not explicitly defined. Consequently, aspects such as multi-tenancy, access control, memory policies, and asynchronous processing required deliberate architectural design and verification. We observe a shift in engineering effort from boilerplate implementation toward constraint specification and enforcement auditing. And we identify recurring architectural “non-delegation zones” where conversational code generation remains insufficient for production reliability.
\end{abstract}



\maketitle

\section{Introduction}

Large language models are reshaping software development through AI-assisted programming and practices such as vibe coding, where developers guide code generation through natural-language interaction with models \cite{sarkar2025vibe}. These tools can accelerate prototyping and implementation. However, they also shift developer effort toward reviewing, validating, and integrating generated code \cite{chang2025codingaireflectionindustrial}. Some developers observe productivity gains, while others report challenges related to reliability, integration, and workflow alignment between generated code and project needs \cite{chen2026beyond,ramler2024industrial,bakajac2025impact}. In conversational development workflows, models generate scaffolding and feature implementations, while developers iteratively refine the output \cite{chang2025codingaireflectionindustrial,sarkar2025vibe}. When architectural constraints are unclear, this interaction can become reactive and exploratory, producing fragile structures or technical debt \cite{waseem2025vibe}. Prior research also shows that prompting strategies and contextual guidance influence the correctness and quality of generated code \cite{yuen2025prompting,gama2025can}.

However, most existing studies have focused on developer productivity, prompting strategies, or perceptions of AI coding assistants \cite{chen2026beyond,ramler2024industrial,SERGEYUK2025107610}. Fewer studies examine how conversational code generation behaves in deployable systems, where architectural constraints such as tenant isolation, access control, retrieval scoping, and asynchronous processing must be enforced.

This paper addresses this gap through an experience report of two deployable AI systems developed using contextual vibe coding under explicit architectural constraints. The first system is a multi-project agent learning platform with structured memory and strict tenant isolation. The second is a file-aware academic RAG system with citation tracing and role-based access control. Both systems were implemented under predefined non-functional requirements, including isolation, asynchronous execution, and controlled retrieval pipelines.

Our \textbf{contributions} are threefold. First, we report practical lessons from building two production-oriented AI systems using conversational code generation. Second, we describe validation practices used to confirm architectural properties such as tenant isolation, role separation, retrieval scoping, and asynchronous processing. Third, we identify recurring architectural tasks that could not be delegated to AI-generated scaffolding and required manual engineering decisions.
\section{Study Design}

The goal was to build deployable systems rather than short-lived prototypes. From the beginning, architectural concerns such as tenant isolation, access control, background processing, and monitoring were treated as core requirements. These constraints reflect challenges reported in prior work when applying AI-assisted development in real software workflows \cite{SERGEYUK2025107610,chang2025codingaireflectionindustrial}.

The first system is a multi-project agent learning platform that uses structured memory and strict tenant isolation. The second is a file-aware academic retrieval-augmented generation (RAG) system that supports document-grounded responses with citation tracing and role-based access control. Both systems required persistent storage, controlled access to data and model outputs, and deployable infrastructure.

\textbf{Requirement Elicitation}: Requirements were defined through structured discussions between developers and stakeholders. These discussions focused on identifying the operational goals of each system and the constraints required for deployable use. The outcomes were documented as functional and non-functional requirements before implementation began. Functional requirements included conversational interaction, document retrieval, role-based access control, administrative dashboards, agent memory updates, and activity logging. Non-functional requirements emphasized system-level properties required for deployment, including tenant isolation, project-scoped data access, asynchronous execution of long-running tasks, and reproducible deployment environments. 
These requirements guided architectural decisions such as database structure and background task orchestration, and later provided reference points for validating whether the systems satisfied the intended constraints.

\textbf{Development approach}: Development followed a conversational workflow often referred to as vibe coding \cite{chang2025codingaireflectionindustrial,sarkar2025vibe}. Developers prompted a language model with functional requirements, architectural constraints, and code context. The model generated candidate implementations, including API routes, database models, and utility functions. In practice, generative tools were effective for routine tasks such as endpoint scaffolding, serialization utilities, and interface components. However, architectural concerns were rarely preserved in generated code. Tasks involving multi-tenant data separation, authentication flows, database relationship design, and orchestration of asynchronous processing pipelines consistently required manual correction and redesign \cite{ramler2024industrial}. Development therefore combined rapid code generation with direct human oversight of the architectural structure.

\textbf{Sources of evidence}: The analysis in this paper is retrospective. Evidence was derived from development artifacts produced during the implementation of the two systems. These artifacts included commit histories, issue discussions, prompt iterations used during AI-assisted development, deployment logs, and internal implementation notes. The analysis focused on identifying recurring patterns in the division of labor between developers and generative tools, and examining how conversational code generation influenced architectural decisions and integration practices.

\textbf{Validation strategy}: Validation examined whether the implemented systems satisfied the architectural properties defined during requirement elicitation. These properties included tenant isolation, role-based access control, retrieval scoping, and reliable asynchronous execution. Evidence for validation came from three sources. Structured manual testing simulated user interactions under different roles and project contexts. Code inspection verified that database queries, authentication checks, and retrieval logic enforced the intended constraints. Runtime logs and background task monitoring were used to examine the behavior of asynchronous pipelines during concurrent operations.

\section{Case Design and Implementation}

This section presents two production-oriented systems developed using contextual vibe coding under explicit architectural constraints. The cases were selected to illustrate how conversational code generation behaves in deployable environments where properties such as tenant isolation, access control, and asynchronous processing must be enforced.
\begin{figure*}
\centering
\includegraphics[width=0.7\linewidth]{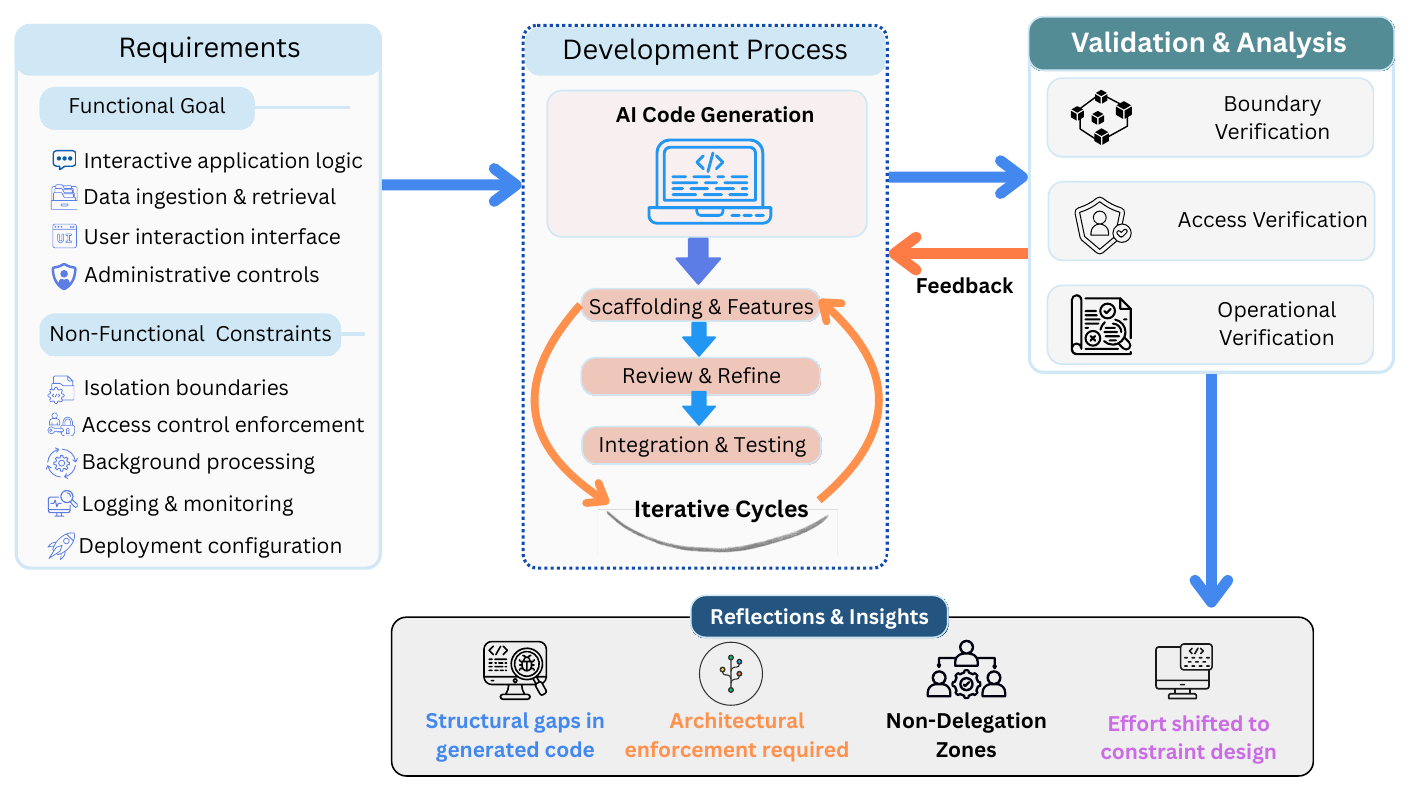}
\caption{Overview of contextual vibe coding in production-oriented systems.}
\label{fig:contextual-vibe-coding}
\end{figure*}

\subsection{Case 1: Learning Platform}

\textbf{Context}:
The first case study is a collaborative agent learning platform designed for organizations that manage shared knowledge through conversational interaction with AI assistants. Project teams upload documents, interact with an agent through chat, and refine the agent’s knowledge collaboratively. The main goal is to support shared knowledge development while maintaining strict separation between projects within a single deployment. Organizations can create multiple projects, each with its own users, documents, and agent. The agent learns from uploaded files, selected group messages that mention the agent, and administrator feedback. When conflicting information appears, the system creates a review ticket and pauses memory updates until an administrator resolves the issue. After resolution, the agent updates its memory following the structured model proposed in Hindsight \cite{latimer2025hindsight}.

\textbf{Requirements}:
Functional requirements included project creation, document upload, conversational interaction with the agent, controlled learning from selected interactions, and an administrator-driven conflict resolution workflow. Non-functional requirements emphasized tenant isolation across organizations and projects, project-scoped storage of knowledge and memory, role-based access control for administrative actions, asynchronous execution of embedding and memory updates, and operational logging. These requirements guided the architectural design and validation.

\textbf{Architecture}:
To enforce project-level isolation, the system adopts a layered architecture that separates interactive operations from background processing. The application layer manages chat interactions, document uploads, and administrative actions. A persistence layer stores organizations, projects, users, and activity logs. Each project is associated with an isolated memory bank to prevent knowledge leakage between projects. A retrieval layer connects user queries to project-specific memory and external language model services. Long-running tasks such as document ingestion and memory updates run asynchronously in a background task subsystem, ensuring that user interactions remain responsive. The platform is implemented with a Python web backend, relational persistence, and background worker queues. Conversational responses are generated through the OpenAI API using the GPT-4.1 model.

\textbf{Use of Generative AI in Development}:
Generative AI was used extensively to scaffold API routes, database models, utilities, and frontend interface components. For routine implementation tasks, the generated code often worked after minor refinement. However, the generated backend code frequently ignored architectural constraints. Early routes omitted project-level filtering in database queries, creating the risk of cross-project data access. Initial memory update routines also attempted synchronous execution despite the requirement for asynchronous processing. These issues required manual correction. Developers added explicit project identifier checks and moved memory updates to background workers. In practice, generative tools significantly accelerated routine implementation but did not reliably preserve architectural constraints.

\textbf{Validation}:
Validation focused on confirming the architectural properties defined in the requirements. Tenant isolation was tested by creating multiple organizations and attempting cross-project interactions through both the interface and API endpoints. These tests verified that database queries consistently enforced project identifiers. Role-based access control was validated by executing administrative actions under different user roles and confirming that restricted operations were rejected. Memory isolation was verified by ensuring that each project mapped to a distinct memory bank and that retrieval routines remained scoped to the correct project. Asynchronous processing was evaluated by triggering concurrent embedding and memory update tasks while monitoring background queues and system logs. These checks confirmed that the system satisfied its architectural constraints while revealing several cases where AI-generated scaffolding required manual correction.

\subsection{Case 2: Academic RAG System}

\textbf{Context}:
The second case study is an academic retrieval-augmented generation (RAG) platform designed for students and researchers working with document collections. Its goal is to allow users to query document repositories and receive answers grounded in uploaded sources with explicit citations. Users interact with the system through a chat interface. Queries trigger the retrieval of relevant document segments that are passed to the language model as contextual input for response generation. When the system produces low-confidence responses, the query is automatically added to an administrator review queue for further inspection \cite{hasan2025engineering,chen2024exploration}.

\textbf{Requirements}:
Functional requirements included document upload, query-based interaction with document collections, retrieval-grounded response generation, citation production, and administrative review of uncertain responses. Non-functional requirements emphasized document-level access control, role separation between student and administrator dashboards, citation traceability, evaluation logging, and non-blocking document ingestion pipelines. These requirements guided both architectural design and validation.

\textbf{Architecture}:
The system follows a layered RAG architecture designed to maintain reliable retrieval and scalable ingestion. The application layer manages user queries and administrative operations. A document processing pipeline performs chunking and embedding generation, while a retrieval component identifies relevant document segments used as context for response generation. Retrieval metadata is stored to support citation tracing and later evaluation of generated answers. Document ingestion and embedding generation run asynchronously to prevent large uploads from blocking user queries. The implementation uses a Python web backend with relational storage, vector retrieval infrastructure, and background worker queues. Responses are generated through the OpenAI API using the GPT-4.1 model.

\textbf{Use of Generative AI in Development}:
Generative AI assisted with implementing ingestion pipelines, retrieval utilities, and citation formatting logic. For many routine components, the generated code provided a useful starting point that developers refined during integration. However, generated implementations frequently ignored system-level constraints. Early versions produced citations without verifying alignment between answers and retrieved document segments. Initial ingestion scripts also attempted synchronous embedding generation, which caused long response times during large uploads. Developers corrected these issues by introducing citation alignment checks and moving embedding generation to background workers. This experience again showed that AI-generated code favors local functionality and requires manual oversight to preserve architectural constraints.

\textbf{Validation}:
Validation focused on confirming that responses remained grounded in authorized document sources. Retrieval scoping was tested by uploading documents with different access permissions and executing queries under multiple user accounts. These checks verified that only authorized documents were retrieved. Citation correctness was evaluated by comparing generated citations with stored retrieval metadata to confirm alignment between answers and retrieved segments. Role separation was validated by performing administrative actions under different accounts. Finally, ingestion reliability was assessed by uploading large documents and verifying that embedding jobs executed asynchronously without blocking user queries. 
\section{Lessons Learned}

Vibe coding accelerated early scaffolding and feature implementation in both systems and allowed functional prototypes to appear quickly. However, stabilizing the systems required deliberate architectural decisions and repeated correction of generated code. The following lessons summarize recurring patterns observed during development and validation.

\textbf{Architectural Constraints Must Be Explicit}:
Generated implementations reflected the level of architectural detail provided in the prompts. Vague prompts often produced locally correct but structurally incomplete solutions. Isolation rules, role checks, and background processing logic were usually absent unless explicitly specified \cite{yuen2025prompting,waseem2025vibe}. Generated code also tended to optimize for minimal functionality and lacked awareness of broader system context, a limitation observed in prior studies of AI coding assistants \cite{SERGEYUK2025107610,chen2026beyond}. As a result, architectural boundaries such as tenant isolation and access control required systematic auditing and manual enforcement.

\textbf{Policies and Evaluation Require Human Design}:
Generative tools could scaffold interfaces for memory updates or citation logging, but they could not infer the policies governing these processes. In the agent platform, developers had to determine when conversational information should become persistent knowledge \cite{latimer2025hindsight}. In the RAG system, citation tracing and answer validation required explicit evaluation criteria and logging mechanisms \cite{imtiyaz2025assessing}. These decisions depended on domain context and therefore remained human-driven.

\textbf{Infrastructure Decisions Cannot Be Deferred}:
Early prototypes executed inference and embedding tasks synchronously. This approach quickly led to instability during repeated use. Background workers, caching, and monitoring were required to prevent blocking operations and performance degradation \cite{kurra2025generative,hasan2025engineering}. These observations show that infrastructure design must be addressed early when building deployable AI systems.

\textbf{Engineering effort shifts toward constraints and validation}:
Vibe coding reduced effort for routine scaffolding and boilerplate generation. However, engineering work shifted toward architectural specification, constraint enforcement, validation, and operational monitoring. Isolation auditing, citation validation, and background job inspection became recurring development activities. This pattern aligns with prior observations that AI-assisted coding shifts effort toward review and verification rather than eliminating it \cite{ramler2024industrial,SERGEYUK2025107610,chang2025codingaireflectionindustrial}. Table~\ref{tab:effort} summarizes this redistribution of engineering effort.

\begin{table}[h]
    \caption{Observed Shift in Engineering Effort}
    \label{tab:effort}
    \begin{tabular}{p{4cm} p{4cm}}
        \toprule
        Reduced Effort & Increased Effort \\
        \midrule
        Boilerplate writing & Architecture design \\
        CRUD scaffolding & Isolation auditing \\
        Basic routing & Policy specification \\
        UI template creation & Validation and monitoring \\
        \bottomrule
        \end{tabular}
\end{table}
\section{Related Work}

Prior research examines AI-assisted coding, conversational development workflows, and architectures for deployable language-model systems. This section situates our study within these research streams and highlights how our work extends them.

\subsection{AI-Assisted Coding and Productivity}

Empirical studies examine how language model–based coding assistants influence developer workflows and productivity \cite{chang2025codingaireflectionindustrial,SERGEYUK2025107610,ramler2024industrial}. These tools accelerate tasks such as scaffolding, code completion, and boilerplate generation, enabling rapid prototyping. However, several studies report that engineering effort shifts toward validation, debugging, and architectural reasoning rather than disappearing \cite{chang2025codingaireflectionindustrial,ramler2024industrial}.

Industry and survey reports also highlight reliability issues, limited project context, and integration challenges in real development workflows \cite{SERGEYUK2025107610}. In contrast, our work examines how AI-generated implementations behave under architectural constraints required for deployable systems, including tenant isolation, access control, and asynchronous processing.

\subsection{Vibe Coding and Context Engineering}

Vibe coding describes a development style in which developers guide implementation through conversational interaction with language models \cite{sarkar2025vibe}. Prompts and iterative feedback replace direct code writing as the primary mechanism for shaping system behavior. Prior work places this approach along a spectrum from suggestion-based completion tools to more autonomous development workflows \cite{chang2025codingaireflectionindustrial}.

Recent research highlights a flow–debt tradeoff in conversational development. Rapid code generation can accelerate early implementation but may introduce architectural fragility when constraints are underspecified \cite{waseem2025vibe}. Studies on prompt and context engineering also show that structured prompts and explicit goals strongly influence the correctness and maintainability of generated code \cite{yuen2025prompting}.

However, this work rarely examines how conversational code generation behaves when strict architectural constraints must be enforced. Our study extends this line of research by analyzing contextual vibe coding in systems requiring multi-tenancy, role-based access control, and structured validation.

\subsection{RAG Systems and Agent Memory}

Retrieval-augmented generation (RAG) architectures improve reliability by grounding model outputs in external knowledge sources. Production deployments typically involve document chunking, embedding pipelines, vector retrieval, caching, and evaluation logging \cite{kurra2025generative}. These systems must also enforce access boundaries to prevent unauthorized data exposure.

Structured agent memory architectures extend this approach by introducing persistent stores for experiences, facts, and belief updates. Frameworks such as Hindsight separate these components to support traceable knowledge updates and long-term reasoning \cite{latimer2025hindsight}. Such systems require explicit policies governing memory updates and controlled retrieval.

Prior work mainly examines architectural design patterns for these systems. Our work differs by analyzing how such architectures are developed using conversational code generation and by identifying architectural elements that remain resistant to automated implementation in production environments.
\section{Threats to Validity}

This study has several limitations that should be considered when interpreting its findings.

First, the analysis is based on two systems developed by a small team with prior full-stack and infrastructure experience. Development practices and architectural decisions may vary in larger organizations, among teams with different levels of expertise, or across projects using alternative technology stacks or model providers. Consequently, the findings should be interpreted as experience-based insights rather than broadly generalizable conclusions.

Second, the study relies on retrospective analysis of development artifacts, including commit histories, implementation notes, system logs, and prompt iterations. Such retrospective interpretation may introduce bias because the findings are reconstructed after development rather than captured in real time.

Third, the authors of this paper were directly involved in the design and implementation of the two systems, which introduces a potential risk of author bias. To mitigate this risk, the analysis was grounded in observable artifacts, including source code changes, runtime logs, and documented prompt interactions, rather than relying solely on subjective recollection.

Finally, validation was based on structured manual testing, code inspection, and runtime log analysis. We did not conduct controlled experiments or quantitative assessments of productivity, code quality, or defect rates. Future work could strengthen these findings through controlled studies, comparative evaluations, and analyses involving larger and more diverse teams.
\section{Conclusion}

This paper presented an experience report on building two deployable production ready systems using contextual vibe coding under architectural constraints, such as tenant isolation, role-based access control, asynchronous processing, and structured memory management. Our observations show that vibe coding accelerates scaffolding and routine implementation tasks such as API routes, utility functions, data models, and user interface components. However, it does not replace architectural design. Critical system properties such as isolation boundaries, access control enforcement, memory update policies, evaluation pipelines, and deployment configuration required manual engineering decisions and careful auditing of generated code.

In both cases, engineering effort shifted from writing boilerplate toward defining architectural constraints, verifying generated implementations, and stabilizing infrastructure. These findings suggest that AI-assisted development is most effective when used as a development tool for rapid implementation within clearly specified architectural boundaries rather than as a replacement for system-level design.

Future research should further investigate development practices, validation strategies, and architectural patterns that better support production-oriented use of conversational code generation in larger teams and higher-scale deployments.

\bibliographystyle{ACM-Reference-Format}
\bibliography{Ref}

\end{document}